\documentclass[aps,pra,epsfigure,twocolumn,longbibliography]{revtex4-1}
\usepackage{dcolumn}    
\usepackage{bm} 
\usepackage{graphicx}
\usepackage{amsmath}    
\usepackage{enumitem}
\usepackage{latexsym}
\usepackage{amsfonts}   
\usepackage{amssymb}
\usepackage{array}      
\usepackage{epsfig}
\usepackage{txfonts}
\usepackage{color}
\usepackage[colorlinks=true,linkcolor=blue,urlcolor=blue,citecolor=blue,pdfusetitle]{hyperref}
\usepackage{hyperref}

\newcommand{\ket}[1]{\left\vert#1\right\rangle}

\newcommand{\blah}{blah\\blah\\blah\\blah.}
\newcommand{\eq}{Eq.~}
\newcommand{\eqs}{Eqs.~}
\newcommand{\fig}{Fig.~}

\newcommand{\cf} {cf.~}

\newcommand{\rref} {Ref.~}

\catcode`\|=\active \def|{
\fontencoding{T1}\selectfont\symbol{124}\fontencoding{\encodingdefault}}

\newcommand{\nobracket}{}
\newcommand{\nosymbol}{}

\newcommand{\tmop}[1]{\ensuremath{\operatorname{#1}}}
\newcommand{\tmtextbf}[1]{{\bfseries{#1}}}

\begin{document}
\title{System-environment correlations and Markovian embedding of quantum non-Markovian dynamics}
       
\author{Steve Campbell,$^{1}$ Francesco Ciccarello,$^{3,4}$ G. Massimo Palma,$^{3,4}$ and Bassano Vacchini$^{2,1}$}
\affiliation{\mbox{$^1$Istituto Nazionale di Fisica Nucleare, Sezione di Milano, via Celoria 16, 20133 Milan, Italy}\\
\mbox{$^{2}$Dipartimento di Fisica ``Aldo Pontremoli", Universit{\`a} degli Studi di Milano, via Celoria 16, 20133 Milan, Italy}\\
\mbox{$^{3}$Dipartimento di Fisica e Chimica, Universit$\grave{a}$  degli Studi di Palermo, via Archirafi 36, I-90123 Palermo, Italy}\\
$^{4}$NEST, Istituto Nanoscienze-CNR}

\begin{abstract}
We study the dynamics of a quantum system whose interaction with an environment is described by a collision model, i.e. the open dynamics is modelled through sequences of unitary interactions between the system and the individual constituents of the environment, termed ``ancillas", which are subsequently traced out. In this setting non-Markovianity is introduced by allowing for additional unitary interactions between the ancillas. For this model, we identify the relevant system-environment correlations that lead to a non-Markovian evolution. Through an equivalent picture of the open dynamics, we introduce the notion of ``memory depth" where these correlations are established between the system and a suitably sized memory rendering the overall system+memory evolution Markovian. We extend our analysis to show that while most system-environment correlations are irrelevant for the dynamical characterization of the process, they generally play an important role in the thermodynamic description. Finally, we show that under an energy-preserving system-environment interaction, a non-monotonic time behaviour of the heat flux serves as an indicator of non-Markovian behaviour.
\end{abstract}
\date{\today}
\maketitle

\section{Introduction}
The inherent fragility of pure quantum states necessitates that we develop a firm understanding of how these systems interact with their environments. If the system under scrutiny is weakly coupled to the environment and/or the latter has negligible correlation time, the ensuing dynamics can be well approximated using Markovian master equations~\cite{OpenQS}. These typically rely on the assumption that the environment remains unaffected by its coupling to the system thereby yielding negligible system-reservoir correlations. Such an approximation generally needs to be relaxed in the description of non-Markovian dynamics. Equally important with understanding the underlying mechanisms that govern the dynamics of an open quantum system is assessing their thermodynamic principles~\cite{JohnReview}. Advances in this regard have been made for both Markovian~\cite{EspositoPRX, Spohn, DeffnerPRL2011, PaternostroArXiv, PaternostroPRL2017} and, to a lesser extent, non-Markovian dynamics~\cite{ManiscalcoSciRep, MarcantoniSciRep,  GiacomoPRA2016a, GiacomoPRA2016b,  PatiPRA, Popovic2018}. 

Of course, not all dynamics fall within the memoryless environment paradigm and as such there has been rapid developments in tools to model quantum non-Markovian dynamics~\cite{Breuer2016a, CampbellPRA2012, BreuerPRL, XueArXiv, XueIEEE, TamascelliPRL2018}. Within the plethora of approaches, collision models (CMs) stand out. In its simplest memoryless version, a CM assumes that the reservoir consists of a large number of initially uncorrelated subunits or ``{\it ancillas}" with which the open system collides one at a time \cite{rau,brun,scarani2002,BuzekPRA}. Suitable modifications endow this basic CM with memory, making it an advantageous tool to tackle quantum non-Markovian dynamics \cite{MassimoPRL,aspuru2012,rybar2012,CiccarelloPRA2013,SantosPRA2012,StrunzPRA2016, lorenzo2016,lorenzo2017a,sabrinaCM, BodorPRA, vegaCM,SantosPRA2017, JinArXiv, LoFranco2018}. In addition, CMs have found application in quantum optics~\cite{grimsmo2015,pichler2016,milburn-combes,FrancescoQMQM}, quantum gravity \cite{KTM2014,altamirano2017}, quantum control \cite{datta,kempf2016,Strunz2018,BarisArXiv} and quantum thermodynamics~\cite{LorenzoPRA2015, ChenPRA2017, PezzuttoNJP, RuariPRL, Diosi, RennerArXiv, PereiraPRE, LiPRE, BarisPRA2017, BarraSciRep, UzdinNJP, HorowitzNJP, HorowitzPRE}.

Remarkably, the discrete nature of CMs together with their tractability allow to study how the progressive inclusion of system-reservoir correlations introduces non-Markovianity in the open dynamics. Along this line, McCloskey and Paternostro \cite{RuariPRA} considered a non-Markovian CM where the system-ancilla collisions are interspersed with nearest-neighbor ancilla-ancilla (AA) collisions, where the latter ones introduce a memory mechanism. They investigated the effect of erasing correlations between $S$ (the system) and the last collided ancilla or between $S$ and the next-to-last ancilla, showing that the latter erasure scheme results in increased non-Markovianity. 

In the CM of \rref\cite{RuariPRA} the dynamical map of $S$, and thus the degree of non-Markovianity, will no longer change if correlations with ancillas prior to the next-to-last one are retained. This is because only nearest-neighbor ancillas collide: once a given ancilla has collided with $S$ and then with the next ancilla, it cannot affect the dynamics of $S$ any more. In fact, only correlations between the system and the portion of environment with which it is currently interacting matter, the {\it size} of such environment's portion being related to the {\it range} of intra-bath interactions. Based on this, one may ask whether, by embedding $S$ into an extended open system  $\mathfrak{S}$ that comprises a small subset of environmental ancillas as well, correlations between such a redefined open system and the remaining environment can now be neglected without affecting the dynamics of $\mathfrak{S}$, and hence of $S$. Such an effective description of a non-Markovian dynamics is sometimes referred to as {\it Markovian embedding} \cite{Siegle2010a,StrunzPRA2016}. According to this picture, it is reasonable to expect that the number of ancillas in $\mathfrak{S}$ should reflect the size of the bath portion whose correlations with $S$ cannot be neglected.

With the main goal of assessing and formalizing the above picture, in this work we explore the connection between system-environment correlations and Markovian embedding. We focus on a non-Markovian quantum CM featuring AA collisions without necessarily restricting to nearest-neighbor AA collisions and with no constraints on the form of the pairwise coupling Hamiltonian ruling both system-ancilla and AA collisions (many CMs including \rref\cite{RuariPRA} focus on partial-SWAP collisions). If $d$ is the range of inter-ancillary collisions (i.e., each ancilla collides with the next $d$ ancillas only) then correlations between $S$ and the last $d$  ancillas it collided with cannot be erased, prior to the intra-environment AA collisions, without affecting the open dynamics of $S$. This observation will motivate us to introduce within this framework the concept of {\it ``memory depth"} as measured by the integer $d$. It will be shown that a Markovian embedding for $S$ holds provided it is incorporated into a composite system $\mathfrak{S}$ featuring in addition as many ancillas as the memory depth $d$. 

While only correlations with a limited portion of the bath are essential to capture the open dynamics, generally all system-environment correlations are instead essential for the thermodynamical properties. In the last part of the work, we will provide evidence of this by considering the decomposition of the system's entropy change in terms of reversible and irreversible entropy production the latter being in particular dependent on system-environment correlations.

The remainder of the paper is organized as follows. In Sec.~\ref{model} we introduce the collision model that we will use. In Sec.~\ref{methods}, we briefly describe how quantities such as the degree of non-Markovianity and correlations will be measured throughout our analysis. In Sec.~\ref{erasure} we explicitly examine the relevance of correlations in a nearest-neighbor CM, recapitulating and extending some known results. In Sec.~\ref{depth} we establish equivalent descriptions of the model in terms of a Markovian embedding and introduce the notion of memory depth. Sec.~\ref{thermodynamics} studies the thermodynamical aspects of the model and in Sec.~\ref{conclusions} we conclude and discuss the generality of our results and their possible application to other non-Markovian models.

\section{System-environment model}
\label{model}
The CM that we consider in this work belongs to the class of non-Markovian CMs where the reservoir's memory mechanism is due to the occurrence of AA collisions \cite{RuariPRA, BarisPRA, ciccarello2013b, lorenzo2016, StrunzPRA2016}, the first of which was introduced in \rref\cite{CiccarelloPRA2013} and stimulated a number of studies on a new class of non-Markovian dynamics \cite{VacchiniPRA2013,vacchini2014,VacchiniPRL,lorenzo2017b,darius2016,darius2017}.

The CM assumes that the reservoir or environment $E$ is made up of a large number of identical ancillas, each called $E_n$. The total state of system and environment is initially factorized , i.e.,
\begin{align}
\rho_{SE}(0) = \rho_S(0) \otimes \rho_{E_1}\otimes \rho_{E_2}\otimes\,...\,,\label{initial}
\end{align}
the initial correlations between $S$ and any ancilla as well as between any two ancillas thus being zero.

\begin{figure}[t]
\includegraphics[width=0.9\columnwidth]{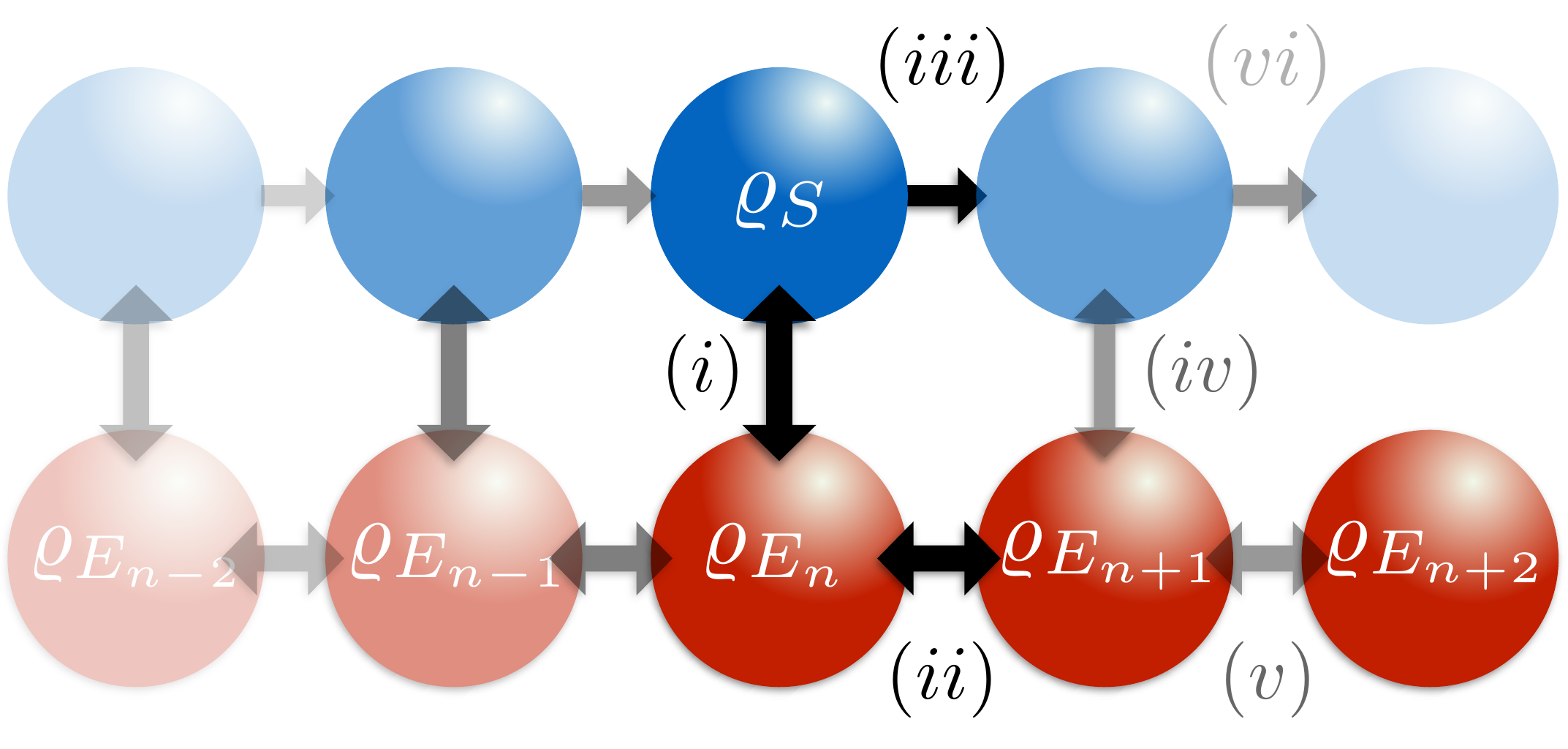}
\caption{{\bf (a)} Schematic of the collision model for nearest-neighbor AA collisions (i.e, $d=1$). In the $n$-th step of the dynamics $(i)$, $S$ collides with $E_{n}$ and next $(ii)$ $E_{n}$ collides with $E_{n+1}$. The system $S$ then moves forward $(iii).$ At the $(n{+}1)$-th step $(iv)$, $S$ collides with $E_{n+1}$ and $(v)$ $E_{n+1}$ collides with $E_{n+2}$. For the case of $d$-range AA collisions, after $S$ has interacted with the ancillas $E_n,\dots E_{n+d-1}$, AA collisions up to range $d$ take place between the ancillas $E_n, \dots, E_{n+d}$.}
\label{fig1}
\end{figure}

The dynamics proceeds through pairwise interaction processes or ``collisions"  between $S$ and a portion of the environment, $\{E_n,\dots E_{n+d-1} \}$. In what follows we assume these collisions to happen successively, however our results hold for simultaneous interactions. These are followed by intra-environment pairwise collisions among $E_n,~E_{n+1},\dots,\,E_{n+d}$, with $d$ the inter-ancillary collision range (later on reinterpreted as memory depth). For $d=1$, corresponding to nearest-neighbor AA collisions, $S$ thus initially collides with $E_{1}$, after which $E_1$ collides with $E_{2}$, then $S$ collides with $E_{2}$ and $E_{2}$ with $E_{3}$ and so on. All collisions are assumed to be unitary. In Fig.~\ref{fig1} we show a schematic of the CM considered. 

If AA collisions are removed, the CM reduces to a fully memoryless one: prior to collision $S$-$E_n$, ancilla $E_n$ is still in its initial state and thus fully uncorrelated with $S$, hence past history cannot affect the dynamics. The occurrence of AA collisions instead endows the environment with memory: if $d=1$ (to fix the ideas) after colliding with $S$, ancilla $E_{n-1}$ undergoes an AA collision with $E_{n}$; thereby $S$ and $E_n$ are already correlated before colliding with each other. An analogous argument holds for $d>1$.

We will assume throughout that $S$ and each ancilla $E_n$ are qubits with free Hamiltonians  $\hat H_S{ =} -\omega_0 \sigma_{Sz}$ and $\hat H_{E_n} {=} -\omega_0 \sigma_{E_n z}$, respectively, with $\sigma_{i=x,y,z}$ being the usual Pauli operators (we set $\hbar=1$). 

The general interaction Hamiltonian describing a collision reads
\begin{equation}
\label{HI}
\hat H_{ij} {=} \!-\!\frac{1}{2} \left(J_x \sigma_{i x} \,\sigma_{j x} {+} J_y \sigma_{i y}\, \sigma_{j y}  {+} J_z \sigma_{i z} \,\sigma_{j z}  \right)\, ,
\end{equation}
(tensor product symbols are not shown) where, for system-ancilla collisions, $i$ and $j$ stand for the system S and an ancilla respectively, while for AA collisions, $i$ and $j$ correspond to two different ancillas with $i\neq j$.

We define a ``step" as the sequence of collisions
\begin{eqnarray}
&\hat U_1=&\prod_{i=1}^{d} e^{- i  \hat H_{S\!E_i}\tau}\,, \\
&\hat U_{n>1}=&\left(\prod_{i=(n-1)d+1}^{nd} e^{- i  \hat H_{S\!E_i}\tau} \right) \left(\prod_{l,m=(n-2)d+1}^{(n-1)d+1}e^{- i  \hat H_{E_l\!E_{m}}\tau}\right)\,, \label{Un}
\end{eqnarray}
for $l\! < \! m$ and where $\tau$ is the collision time and we have allowed for arbitrarily ordered pairwise AA interactions (in what follows for brevity we will denote the action of the AA interactions with $\mathcal{V}_{[\cdot]}$). The evolution operator of the overall system (i.e., $S{+}E$) at step $n$ is thus given by the composition $\hat U_n\cdots\hat U_1$.

For $J_x{=}J_y{=}J_z\!{=}J$ the interaction Hamiltonian (\ref{HI}) is energy-preserving and the corresponding unitary is the partial-SWAP operation $\hat U=\cos(J \tau)\openone-i \sin(J\tau)\hat S$ often considered in the CM literature~\cite{RuariPRA,RuariPRL,PezzuttoNJP, BarisPRA}. Here, angle $J\tau$ (defined such that $0\le J\tau\le \tfrac{\pi}{2}$) measures the strength of AA collisions: these have zero and maximum effect if, respectively, $J\tau{=}0$ and $J\tau{=}\tfrac{\pi}{2}$. In the former case, the model reduces to a fully Markovian one. In the latter case (corresponding to perfect swap), a strongly non-Markovian dynamics occurs instead since $S$ behaves as if it is interacting with the same ancilla at all times~\cite{CiccarelloPRA2013,RuariPRA}.

\section{Non-Markovianity and correlations}
\label{methods}
To quantify the degree of non-Markovianity of the open dynamics of $S$ we will use the criterion underpinning the widely adopted BLP measure~\cite{BreuerPRL} according to which, non-Markovian behavior is associated with back-flow of information from the environment to the system. Occurrence of information back-flow depends on the time behavior of the trace distance between two different initial states of $S$, namely
\begin{equation}
\mathcal{D}_n= \frac{1}{2} \left\| \rho_{Sn}^{(1)} -\rho_{Sn}^{(2)} \right\|_1\,,
\label{tracedist}
\end{equation}
where $\|\cdots\|_1$ is the trace norm \cite{NielsenChuang} while $\rho_{Sn}^{(I)}\!$ is the $n$th-step state of $S$ when it starts in $\rho_{S}^{(I)}\!(0)$. Thus, in the present CM framework time is a discrete variable. Under Markovian dynamics, $\mathcal D$ monotonically decreases with time whatever the initial pair of states $\{\rho_{S}^{(1)}\!(0),\rho_{S}^{(2)}\!(0)\}$. Thereby, a non-monotonic behavior of this quantity for at least one pair of initial states is sufficient to conclude that the governing dynamics is non-Markovian~\cite{BreuerPRL}. However, we remark that this is not a necessary condition as there may be orthogonal pairs resulting in a monotonic decay of the trace distance ${\cal D}_n$, while the dynamics, nevertheless, are non-Markovian. It can be shown \cite{Wissmann2012} that the pair of initial states yielding maximum deviation from a monotonic decay of the trace distance are orthogonal and lie on the boundary of the state space (i.e., the Bloch sphere in the present case of qubits). This underpins our choice of the specific initial pairs of states that we will focus on in our analysis.

A major focus of this work is the relevance of system-environment correlations. Specifically, in terms of the CM introduced in Section \ref{model}, we will be concerned with bipartite correlations between $S$ and a single environmental ancilla, $E_i$. These can be quantified through the quantum mutual information \cite{NielsenChuang} for the state $\rho_{SE_{i}}$
\begin{equation}
\label{mutualinfo}
\mathcal{I}_{SE_{i}}(\rho_{SE_{i}}) =S(\rho_S) + S(\rho_{E_i}) - S(\rho_{SE_{i}}) ,
\end{equation}
where $S(\rho)=-\text{Tr}\left[ \rho \log \rho \right]$ is the von Neumann entropy \cite{NielsenChuang} while $\rho_S={\rm Tr}_{E_i}\rho_{SE_{i}}$ is the reduced state of $S$ ($\rho_{E_i}$ is defined analogously). It is widely accepted that the mutual information captures the full amount of correlations, both classical and quantum. For product states $\rho_{SE_{i}}=\rho_S\otimes\rho_{E_i}$ the mutual information always vanishes. In the general case, however, $\rho_S\otimes\rho_{E_i}$ does not equal $\rho_{SE_{i}}$ and the mutual information is non-zero. Based on this, we see that in the general case the replacement $\rho_{SE_{i}}\rightarrow \rho_S\otimes\rho_{E_i}$ (which is a quantum map) extracts the uncorrelated part of the joint state $\rho_{SE_{i}}$ and, as such, erases the $S$-$E_i$ correlations altogether.

At this point we highlight an important observation. It is worth recalling that the knowledge of $\rho_S\!=\!{\rm Tr}_{E_i}\rho_{SE_{i}}$ is sufficient to determine the final state of $S$, $\rho'_S$, after a quantum map on $S$ (unitary or not) ${\cal M}_S\otimes{\cal I}_{E_i}$ is applied on $\rho_{SE_{i}}$. In such cases, thereby, {\it if} one is interested in the state of $S$ only, erasing correlations, i.e., applying the map $\rho_{SE_{i}}\rightarrow \rho_{S}\otimes \rho_{E_i}$, has no effect.

\section{Relevant Correlations: Open Dynamics} 
\label{erasure}
It is instructive for our purposes to see how the last property discussed in the previous section applies to a memoryless CM (namely the CM of Fig.~\ref{fig1} in the absence of AA collisions). Before colliding with $E_n$, $S$ is correlated with all previous ancillas. Yet, only its reduced state right before the collision with $E_n$ is needed to work out the open dynamics. This means that correlations between $S$ and each ancilla can be erased after they have collided without affecting the open dynamics of $S$, which in practice means it is enough to keep track of the state of $S$ only throughout (that is only one qubit).

In line with \rref\cite{RuariPRA}, the approach that we adopt to investigate the importance of correlations is studying how the repeated {\it erasure} 
of system-ancilla correlations affect the open dynamics of $S$, in particular, non-Markovianity of the system's evolution and its thermodynamical properties. Specifically, we consider erasing correlations according to the following three schemes:
\begin{itemize}
\item Scheme {\bf A}: The correlations established after $S$ has collided with $E_n$ are erased {\it before} $E_n$ interacts with $E_{n+1}$. 
\item Scheme {\bf B}: The correlations established after $S$ has collided with $E_n$ are erased only {\it after} $E_n$ has collided with $E_{n+1}$ but {before} $S$ collides with 
$E_{n+1}$.
\item Scheme {\bf C}: The correlations established after $S$ has collided with $E_n$ are retained, like scheme {\bf B}, during the $E_n$-$E_{n+1}$ collision and, at variance with {\bf B}, even during the $S$-$E_{n+1}$ collision. Afterward, they are erased {\it before} $E_{n+1}$ collides with $E_{n+2}$.
\end{itemize}
In practice, the above means that, in addition to the state of $S$, we need to keep track of the state of one environment ancilla (scheme {\bf A}), two ancillas (scheme {\bf B}) and three ancillas (scheme {\bf C}). Following from the last property outlined in Sec.~\ref{methods}, the dynamics for schemes {\bf B} and {\bf C} will necessarily be identical, however we explicitly consider this case as it will be instructive when assessing the thermodynamics of the CM in Sec.~\ref{thermodynamics}. Schemes {\bf A} and {\bf B} are precisely those considered in Ref.~\cite{RuariPRA} where all interactions were a partial-SWAP.

\begin{figure}[t]
{\bf (a)}\\
\includegraphics[width=0.75\columnwidth]{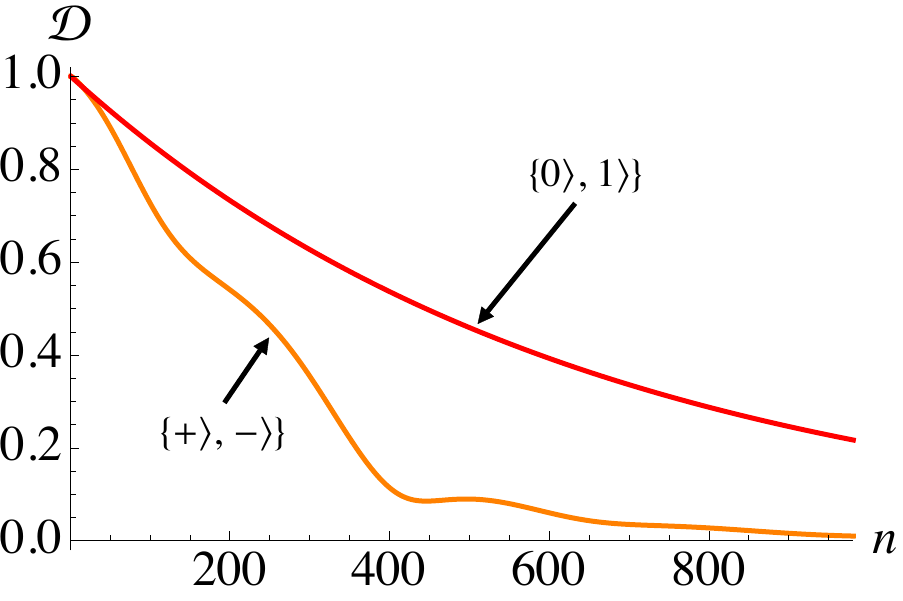}\\ 
{\bf (b)}\\
\includegraphics[width=0.75\columnwidth]{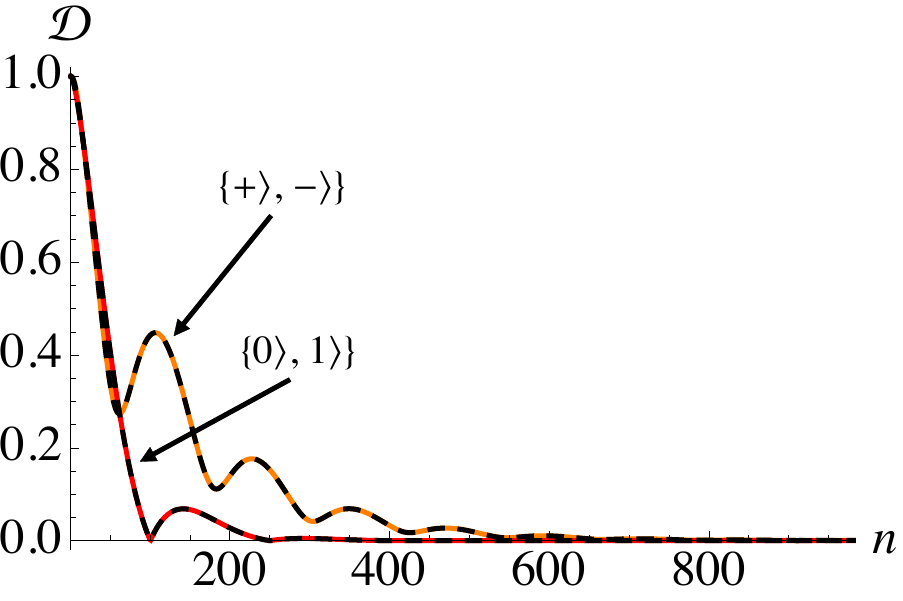}
\caption{{\bf (a)} and {\bf (b)} Trace distance ${\cal D}$ against the number of steps for two representative pairs of initial orthogonal pure states of $S$ taken to be $\{|0\rangle, |1\rangle\}$ and $\{|\pm\rangle\}$ with $|\pm\rangle{=}\left(|0\rangle{\pm}|1\rangle\right)/\sqrt{2}$ under the erasure schemes {\bf A} [panel {\bf (a)}], {\bf B} [solid colors in panel {\bf (b)}] and {\bf C} [panel {\bf (b)}, dashed black]. Note that in panel {\bf (b)} the curves are identical between the two schemes {\bf B} and {\bf C} indicating the irrelevance of the additionally retained correlations. We fix set $J_x\!=\!2J_y\!=\!J_z\!=\!1$ [\cf Eq.~\eqref{HI}] for both system-ancilla and AA collisions and [\cf\eqref{Un}] $J\tau_{SA}{=}0.05$, $J\tau_{AA}{=}0.95\tfrac{\pi}{2}$. In both panels each ancilla is initialized in $|0\rangle_n$.}
\label{fig2}
\end{figure}

\begin{figure}[t]
{\bf (a)}\\
\includegraphics[width=0.75\columnwidth]{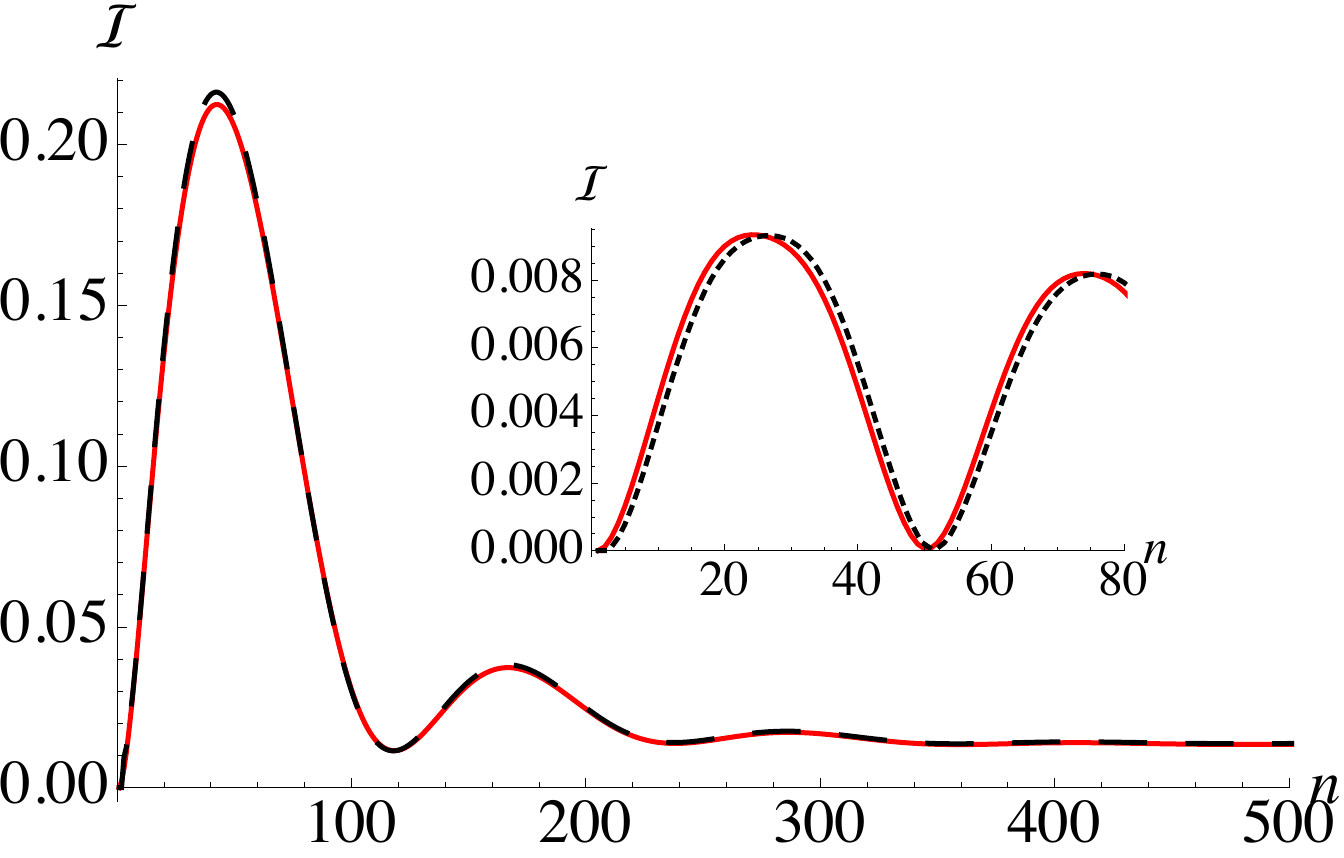}\\ 
{\bf (b)}\\
\includegraphics[width=0.75\columnwidth]{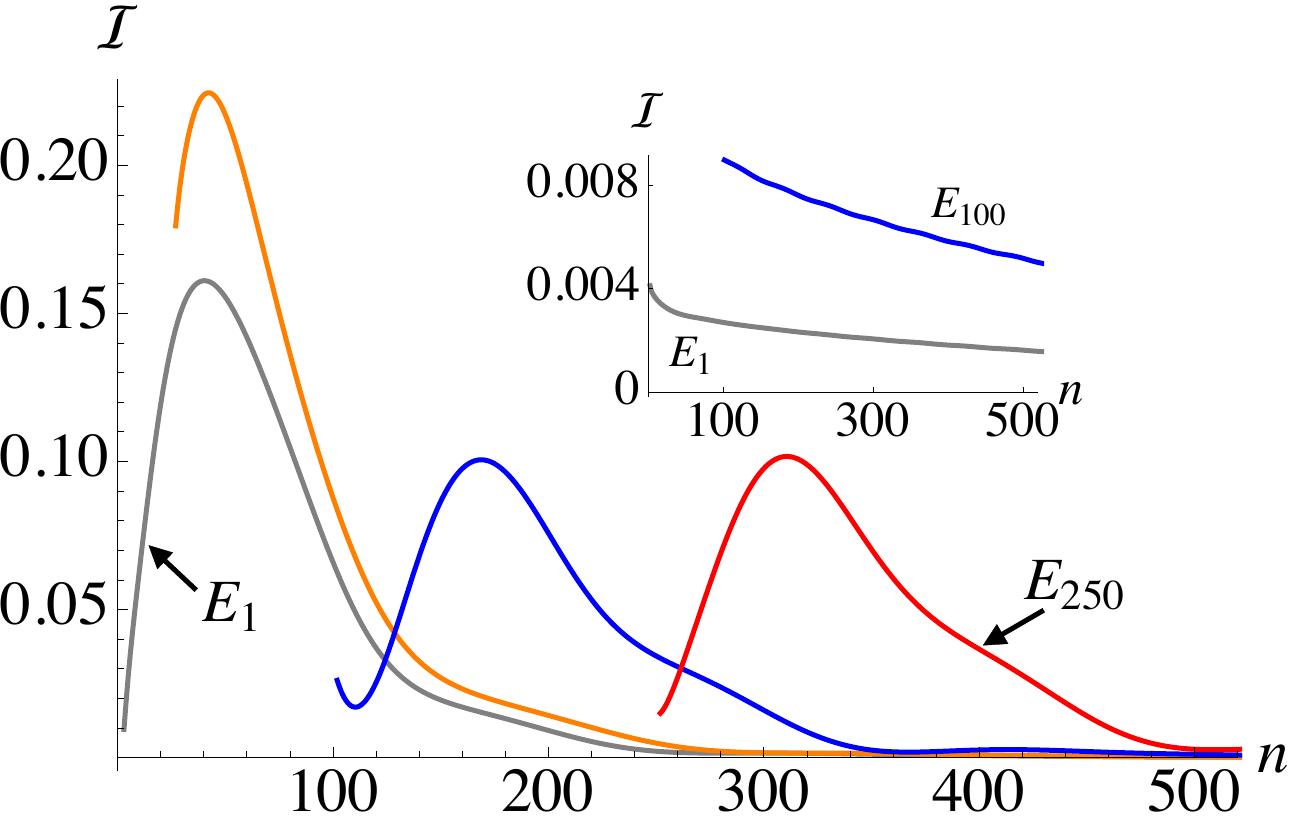}
\caption{{\bf (a)} Quantum mutual information, Eq.~\eqref{mutualinfo} for scheme {\bf C} between $S$ and ${E_n}$ (dashed, black) and ${E_{n-1}}$ (solid, red) for the same parameters as panel Fig.~\ref{fig2} {\bf (b)} and with the system initialized in $\ket{+}$. {\it Inset:} For comparison we also show the same quantities when the partial-SWAP operation is used for all collisions as done in Ref.~\cite{RuariPRA}. {\bf (b)} Quantum mutual information shared between the system and a fixed environment ancilla [curves starting from left to right] $E_1$, $E_{25}$, $E_{100}$, and $E_{250}$, against number of steps, $n$, setting $J_x\!=\!2J_y\!=\!J_z\!=\!1$ and initializing the system in $\ket{+}$. {\it Inset:} As for the main panel except switching off the AA interaction so that the dynamics is fully Markovian. In all plots each ancilla is initialized in its ground state.}
\label{fig3}
\end{figure}

In Fig.~\ref{fig2} {\bf (a)} and {\bf (b)} we examine the behavior of the trace distance, Eq.~\eqref{tracedist}, for different orthogonal pairs of the system's initial state. We have (arbitrarily) assumed all environmental qubits are initialized in their ground states and consider $J_x\!\!=\!\!2J_y\!\!=\!\!J_z$. In panel {\bf (a)} we see that when the correlations are erased before the intra-environment interactions take place, i.e. scheme {\bf A}, the trace distance ${\cal D}(t)$ [\cf\eqref{tracedist}] exhibits a non-trivial, yet monotonically decaying, behavior for both the initial pairs $\{|\pm\rangle\}$ and $\{| 0 \rangle, \ket{1}\}$. However, the less invasive correlations erasure prescribed by scheme {\bf B} leads to an increase of non-Markovian behavior [\cf solid curves in panel {\bf (b)}]. We confirm that the dynamics are unaffected when more environmental ancillas are retained following scheme {\bf C}, as evidenced by the dashed curves in panel {\bf (b)}, the resulting trace distance is invariant and it is easy to confirm that the dynamical state of the system is identical under schemes {\bf B} and {\bf C}. It follows that retaining any more environmental subunits has no effect on the resulting dynamics of the system. One can compare these figures with those of Ref.~\cite{RuariPRA} to assess the dependence of the non-Markovianity on the details of the interactions involved.  In passing, note that the non-Markovianity of the dynamics, as captured by revivals in the trace distance, is significantly diminished by changing the interaction model. This can be understood considering that the partial-SWAP operation used in Ref.~\cite{RuariPRA} as the most `memory-mimicking' interaction one can use and thus leads to the strongest exhibition of non-Markovian features.

It should be stressed that the invariance of the dynamics between Schemes {\bf B} and {\bf C} is notwithstanding the establishment of correlations between $S$ and all previous ancillas, in particular $E_{n-1}$. In Fig.~\ref{fig3} {\bf (a)} we show that there are non-zero correlations shared between the system and ancillas $E_n$ (dashed, black) and $E_{n-1}$ (solid, red) before the $(n\!+\!1)$-th collision. Notice that the $S\!-\!E_{n-1}$ correlations are erased under scheme {\bf B} while they are retained under scheme {\bf C}, and despite this difference the overall evolution of $S$ remains unaffected. This confirms that only certain correlations are important in dictating the dynamical features, and therefore non-Markovian nature, of the system's evolution. In the considered setting, where the environment is restricted to nearest neighbor interactions, the correlations shared between the system, $S$ and environment ancillas $E_{n}$ and $E_{n+1}$ completely characterize the open dynamics. All other correlations, i.e. those shared between $S$ and $E_i$ for $n\!>\!i\!\geq\!1$, are irrelevant. This can be seen from panel {\bf (b)} where we fix an ancilla and assess the MI shared between it and the system during the ensuing dynamics. Clearly there are correlations present long after the ancilla has interacted with the system, but such correlations do not affect the open dynamics. Since the ancilla state no longer changes after all its interactions are over, the change in the MI is entirely due to changes in the state of $S$. This behavior is robust to the particular details defining the interactions, as shown in the insets where we compute the same quantities when all interactions are a partial-SWAP in panel {\bf (a)} and even for the Markovian limit when the AA collisions are switched off in panel {\bf (b)}.

As we will show by means of an analytic argument in Sec.~\ref{depth}, the general features shown in Figs.~\ref{fig2} and \ref{fig3} are generic, see in particular Eqs.~\eqref{eq:l1} and \eqref{eq:scatola2}. Furthermore, in the analytic treatment we also consider interactions beyond nearest-neighbor, involving multiple systems, showing that this can be a natural way to introduce a hierarchy of memory effects and the notion of a memory depth.

\section{Markovian embedding and Memory Depth}
\label{depth}
We next provide a general framework supporting an effective description of the open dynamics that, in particular, better clarifies the relevance of system-environment correlations and, most importantly, shows the effectiveness of a Markovian embedding of the open system under study. Both aspects will be linked to the characteristic range of intra-environment interactions (here embodied by the range of AA collisions $d$), which will be reinterpreted as the memory depth. Such an effective description holds regardless of dimensionality and type of coupling of all the involved particles.

We begin by addressing the $d\!=\!1$ case (nearest-neighbor AA collisions). Let us first introduce a suitable notation. The Hilbert spaces of $S$ and the $n$th environmental ancilla are called $\mathcal{H}_{S}$ and $\mathcal{H}_{E_{n}}$, respectively. We call $\mathcal{U}_{SE_{n}}[\cdot]=\hat U_{SE_n}\cdot\hat U_{SE_n}^\dag$ the unitary map describing the $S$-$E_n$ collision and $\mathcal{V}_{E_{n+1} E_{n}}[\cdot]=\hat U_{E_{n+1},E_n}\cdot\hat U_{E_{n+1},E_n}^\dag$ the unitary map describing the $E_n-E_{n+1}$ AA collision.

Consider the joint Hilbert space of $S$ and $E_n$, $\mathcal{H}_{S} \otimes \mathcal{H}_{E_{n}}$, and let 
$\mathcal{T} ( \mathcal{H}_{S} \otimes \mathcal{H}_{E_{n}})$ be the set of all physical states therein.
We will consider a collection of completely positive trace preserving (CPTP) maps, i.e. maps which ensure the states are physical throughout the evolution~\cite{OpenQS, NielsenChuang}, parametrized by the discrete index $n$, 
\begin{eqnarray}
  \Phi_{n} : \mathcal{T} ( \mathcal{H}_{S} \otimes \mathcal{H}_{E_{n-1}} ) &
  \rightarrow & \mathcal{T} ( \mathcal{H}_{S} \otimes \mathcal{H}_{E_{n}} )
  \nonumber\,,
\end{eqnarray}
which sends states in $\mathcal{T} ( \mathcal{H}_{S} \otimes \mathcal{H}_{E_{n-1}}
)$ to states in $\mathcal{T} ( \mathcal{H}_{S} \otimes \mathcal{H}_{E_{n}} )$, defined as
\begin{align}
\Phi_{1} [w] &= \mathcal{U}_{SE_{1}} [w]\label{Phi1w}\,,\\
\Phi_{n> 1} [w] & =  {\rm Tr}_{E_{n-1}} \mathcal{U}_{SE_{n}} \circ
\mathcal{V}_{E_{n} E_{n-1}}  [w \otimes \rho_{E_{n}} ]\,   \label{Phinw}
\end{align}
where $w$ stands for an arbitrary state of $S$ and $E_{n-1}$.
Here $\rho_{E_{n}}$ is the initial state of ancilla $E_n$ [recall that the system and ancillas start in the factorized state \eqref{initial}].

We can next introduce a map that returns at step $n$ the joint state of $S$ {\it and} the last ancilla
which $S$ collided with as
\begin{eqnarray}
  \Phi (n) [ w] & = & \Phi_{n} \circ \ldots \circ \Phi_{1} [
  w] \, \label{Phi(n)}
\end{eqnarray}
such that $\rho_{SE_1}(1){=}\Phi (1) [\rho_S{\otimes} \rho_{E_1}]$, $\rho_{SE_2}(2){=}\Phi (2) [\rho_{SE_1}]$ and so on. Map $\Phi(n)$ is manifestly memoryless, according to virtually all Markovianity criteria proposed in the literature \cite{Breuer2016a}, since it results from the composition of CPTP maps and trivially fulfills $\Phi(n)=\Phi(n-m)\circ \Phi(m)$ for any $1\le m< n$.

We also define the dynamical map of $S$ as $\rho_{S} (n) = \Lambda (n) [ \rho_{S}(0) ]$, which returns the evolved state of $S$ at each step $n$ for any given initial state $\rho_{S} (0)$.
Clearly, if $\Phi(n)$ is known then so is $\Lambda (n)$ once a partial trace over the last collided ancilla is taken: 
\begin{eqnarray}
\rho_{S} (n) & = & \Lambda (n) [ \rho_{S}(0) ]= \tmop{Tr}_{E_{n}}   \Phi(n)  [
\rho_{S}(0) \otimes \rho_{E_{1}} ] .  \label{eq:scatola}
\end{eqnarray}
Unlike $\Phi(n)$, the dynamical map $\Lambda(n)$ is, in general, non-Markovian (for instance in the case that AA collisions are full swaps, see end of Section \ref{model}). It is however natural to define this open dynamics as {\it ``first-order-Markovian''}, in the sense that, by embedding $S$ into a larger system $\mathfrak{S}$ that comprises of only {\it one} additional environmental ancilla, such a redefined open system $\mathfrak{S}$ undergoes a fully Markovian dynamics described by map $\Phi(n)$ [\cf\eq\eqref{Phi(n)}]. This terminology is introduced by analogy with the corresponding notion in classical stochastic processes \cite{homc}.

A stronger memory (in the above sense), called {\it ``second-order-Markovian''} accordingly, arises for $d\!=\!2$ in which case AA collisions involving three, rather than two, ancillas occur. As before, the overall dynamics results from the application of maps describing system-ancilla and AA collisions, the latter being now described by a {\it tripartite} unitary map $\mathcal{V}_{E_{n+2} E_{n+1} E_{n}}$. The analogue of $\Phi_n$ [\cf\eqs\eqref{Phi1w}-\eqref{Phinw}] now sends physical states of $ \mathcal{H}_{S} \otimes \mathcal{H}_{E_{n-3}} \otimes \mathcal{H}_{E_{n-2}}$ to states of $\mathcal{H}_{S} \otimes \mathcal{H}_{E_{n-1}} \otimes \mathcal{H}_{E_{n}}$ and is defined as
\begin{align}
\Phi_{1}^{(2)} [w ] &=  \mathcal{U}_{SE_{2}}\mathcal{U}_{SE_{1}} [w]\nonumber\\
\Phi_{n>1}^{(2)} [w ] & =  \tmop{Tr}_{E_{2n-2} E_{2n-3}}
\mathcal{U}_{SE_{2n}} \mathcal{U}_{SE_{2n-1}}  \nonumber\\
&\,\,\,\,\,\,\,\,\circ \mathcal{V}_{E_{2n-1} E_{2n-2} E_{2n-3}}  [
w \otimes \rho_{E_{2n-1}} \otimes \rho_{E_{2n}} ]\,, \nonumber
\end{align}
where $w$ now stands for a state in the tripartite Hilbert space $\mathcal{H}_{S} \otimes \mathcal{H}_{E_{2n-3}}
\otimes \mathcal{H}_{E_{2n-2}}$.

In analogy with \eq\eqref{Phi(n)}, the associated map that returns the joint state of $S$ and the last two collided ancillas is given by
\begin{eqnarray}
  \Phi^{(2)}(n) [ w ] & = & \Phi^{(2)}_{n} \circ \ldots \circ \Phi^{(2)}_{1} [
  w]   \label{eq:nm2}
\end{eqnarray}
such that we obtain $\rho_{SE_1E_2}(1){=}\Phi^{(2)} (1) [\rho_S{\otimes} \rho_{E_1}{\otimes} \rho_{E_2}]$, $\rho_{SE_3E_4}(2){=}\Phi^{(2)} (2) [\rho_{SE_1E_2}]$ and so on.
Like map \eqref{Phi(n)}, this is still fully Markovian again because it results from the composition of CPTP maps. The dynamical map of $S$ is obtained from this as [\cf\eq\eqref{eq:scatola}]
\begin{eqnarray}
  \rho_{S} (n) =\Lambda^{(2)} (n) [ \rho_{S} ] =\tmop{Tr}_{E_{2n} E_{2n-1}}\! \Phi^{(2)}(n) [ \rho_{S}{\otimes}\rho_{E_{1}} {\otimes} \rho_{E_{2}}]\, .\nonumber
\end{eqnarray}
The obtained transformations $\{\Lambda^{(2)} (n)\}$ describe a discrete dynamics which can be naturally termed {\it ``second-order-Markovian''}, since it goes over to a Markovian dynamics by enlarging the description to include two environmental ancillas. This notion can naturally be extended by considering reduced dynamics with an arbitrary memory depth $d$. 

The direct connection between memory depth and range of intra-environmental interactions can easily be seen considering the following equivalent representation for the {\it ``first-''} and {\it ``second-order-Markovian''} maps
\begin{eqnarray}
  \Lambda^{( 1 )} (n) [ \rho_{S} ] & = & \tmop{Tr}_{E_{n} \ldots E_{1}} \!
  \mathcal{U}_{SE_{n}} \ldots  \label{eq:l1}\\
  &  & \ldots \mathcal{U}_{SE_{2}}  \mathcal{V}_{E_{2} E_{1}} 
  \mathcal{U}_{SE_{1}} \left[ \rho_{S} \otimes \bigotimes_{i=1}^{n}
  \rho_{E_{i}} \right] \nonumber\,,
\\
  \Lambda^{(2)} (n) [ \rho_{S} ] & = & \tmop{Tr}_{E_{2n} \ldots E_{1}}
  \mathcal{U}_{SE_{2n}} \mathcal{U}_{SE_{2n-1}}\ldots  \label{eq:3b}\\
  &  & \ldots \mathcal{U}_{SE_{4}} \mathcal{U}_{SE_{3}}  \mathcal{V}_{E_{3} E_{2} E_{1}} 
\mathcal{U}_{SE_{2}}  \mathcal{U}_{S E_{1}}  \left[ \rho_{S} \otimes \bigotimes_{i=1}^{2n}
  \rho_{E_{i}} \right] . \nonumber
\end{eqnarray}
It should be noted that the study of memory effects due to interactions involving higher orders of subunits has been considered by \c{C}akmak {\it et al} in Ref.~\cite{BarisPRA}, where it has been shown that indeed this can lead to a higher degree of non-Markovianity according to recently introduced measures~\cite{BreuerPRL}. 

\subsection{Effective description via Markovian embedding}
\begin{figure}[t]
\begin{flushleft}
{\bf (a)}\\
\includegraphics[width=0.99\columnwidth]{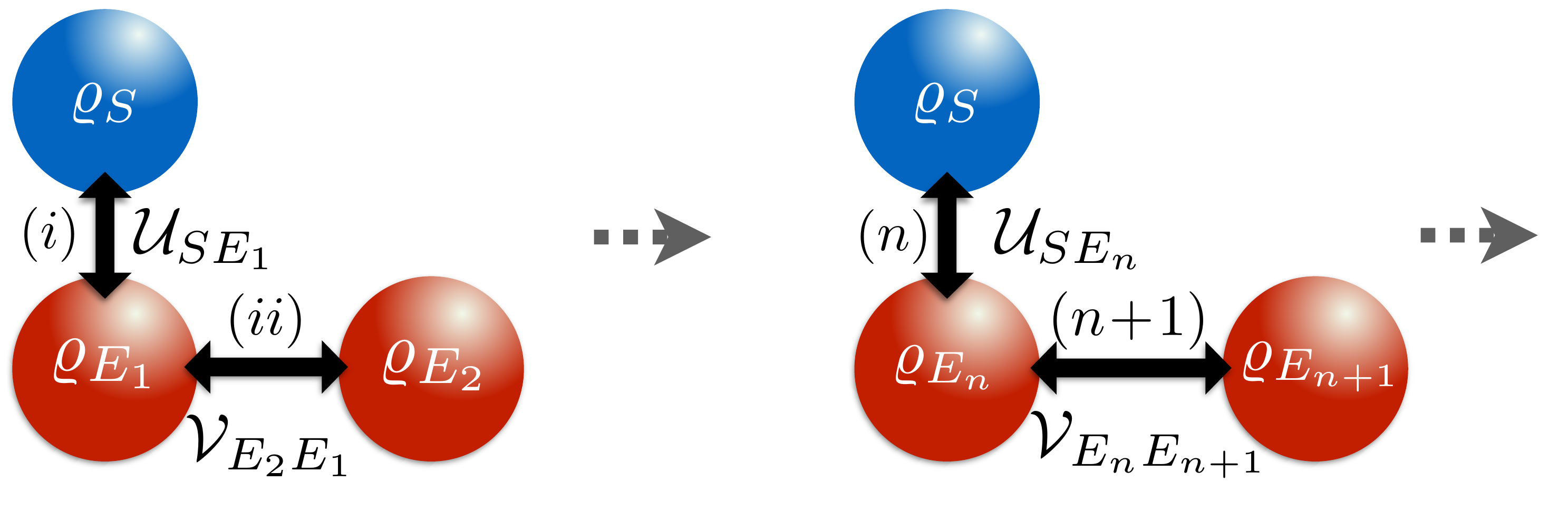}\\
{\bf (b)}\\
\includegraphics[width=0.99\columnwidth]{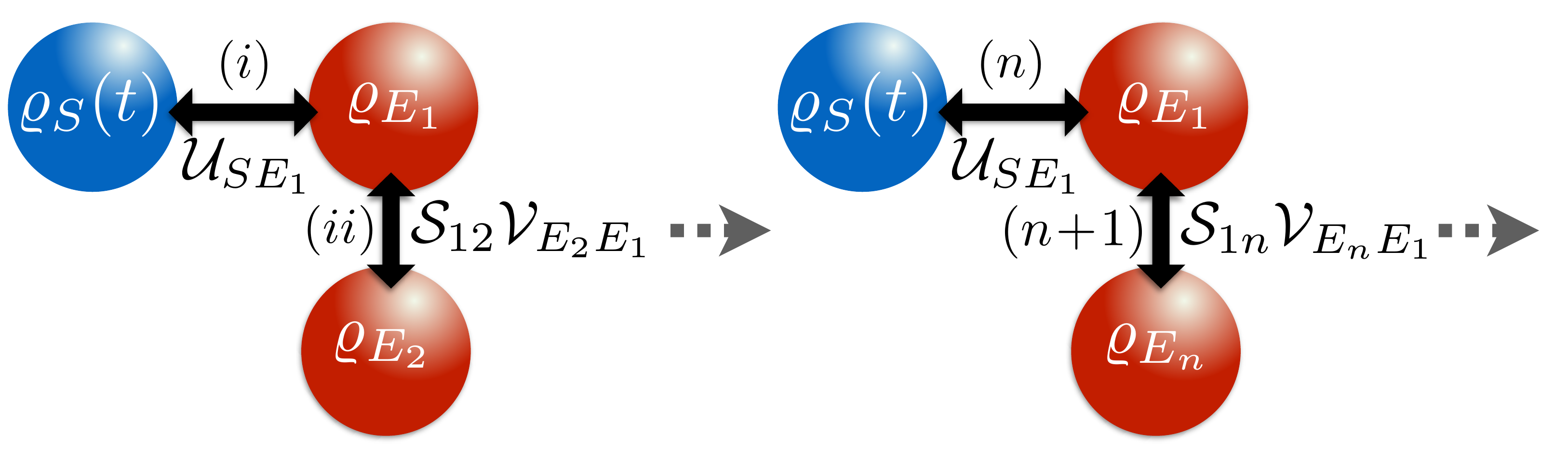}
\end{flushleft}
\caption{Schematics of the different equivalent representations of the open dyanamics of $S$ in the CM of Fig.~\ref{fig1}. {\bf (a)} The system $S$ collides sequentially with the different environmental ancillas, which in turn interact afterward among themselves. {\bf (b)} The system always interacts with the same ancilla (working as a memory), which in turn sequentially collides with the other ancillas.}
\label{FigBassano1}
\end{figure}

Let us first consider the simplest situation where the dynamics can be characterized as {\it ``first-order-Markovian''} such that the evolved system state after $n$ steps is given by Eq.~(\ref{eq:l1}). Due to the fact that the initial state is fully factorized and interactions only take place in a pairwise fashion, the transformation can be arranged in a nested form
\begin{eqnarray}
\rho_{S} (n) & = & \Lambda (n) [ \rho_{S} ] \nonumber\\
  & = & \tmop{Tr}_{E_{n}}\! \mathcal{U}_{SE_{n}} [
  \tmop{Tr}_{E_{n-1}} \mathcal{V}_{E_{n} E_{n-1}}  \mathcal{U}_{SE_{n-1}} [\,
  \ldots   \nonumber
\\
  &  & \tmop{Tr}_{E_{2}}\! \mathcal{V}_{E_{3} E_{2}}  \mathcal{U}_{SE_{2}}
  \nobracket [ \tmop{Tr}_{E_{1}} \!\mathcal{V}_{E_{2} E_{1}} 
  \mathcal{U}_{SE_{1}} [ \rho_{S} {\otimes} \rho_{E_{1}} ] {\otimes} \rho_{E_{2}}
  ] \nobracket \nonumber \\
  &  &  \ldots \,] \nobracket \otimes \rho_{E_{n}} ]\,, \label{eq:scatola2}
\end{eqnarray}
This scheme can be pictorially described through the schematic in Fig.~\ref{FigBassano1} {\bf (a)}. \eq\eqref{eq:scatola2} shows that, in order to fully describe the open dynamics of $S$, it suffices to deal with only {\it three} qubits at each step ($S$ and two ancillas), despite correlations are being generated between $S$ and {\it all} the ancillas which $S$ collided with. In particular, the erasure scheme \tmtextbf{A} (see Section \ref{erasure}) corresponds to replacing in Eq.~(\ref{eq:scatola2}) $\mathcal{U}_{SE_{1}}  [ \rho_{S} \otimes \rho_{E_{1}} ]$ with the product of the first marginal with the initial state of the ancillas, i.e., $\tmop{Tr}_{E_{1}} ( \mathcal{U}_{SE_{1}} [ \rho_{S} \otimes\rho_{E_{1}} ] ) \otimes \rho_{E_{1}}$ (an analogous replacement being made at subsequent steps). Scheme \tmtextbf{B} in Section \ref{erasure}  instead corresponds precisely to the scheme in Eq.~(\ref{eq:scatola2}), while scheme \tmtextbf{C} is recovered by postponing the partial trace by one step.
To rigorously introduce the notion of memory depth, an alternative formulation can be considered, which puts into evidence the existence of a memory ancilla mediating the interaction between system and environment, and connects the memory depth to the size of the memory ancilla. 

Let us first introduce the unitary swap map $\mathcal S_{1,2}$, which exchanges the state of two systems (having the same dimension) as $\mathcal S_{1,2}[\rho_{1} {\otimes} \rho_{2}]{=}\hat S_{1,2}\,\rho_{1} {\otimes} \rho_{2}\hat S_{1,2}{=}\rho_{2} {\otimes} \rho_{1}$ with $\hat S_{1,2}\equiv\hat S_{1,2}^\dag$ the standard swap operator. This fulfills the basic properties
\begin{align}
  \tmop{Tr}_{E_{2} E_{1}} \mathcal{S_{\nosymbol}}_{1,2} [\cdots] &=
  \tmop{Tr}_{E_{2} E_{1}} [\cdots]\,,\label{TrS}
\end{align}
and $ \mathcal{S_{\nosymbol}}_{1,2} \circ\mathcal{V}_{E_{2} E_{1}} =
\mathcal{V}_{E_{1} E_{2}}  \circ\mathcal{S}_{1,2}$. The latter identity, if $E_2$ is replaced with an arbitrary ancilla $E_m$, can be written more generally as
\begin{eqnarray}
  \mathcal{S_{\nosymbol}}_{1,m} \circ \mathcal{V}_{E_{n} E_{m}} =
  \mathcal{V}_{E_{n} E_{1}} \circ  \mathcal{S}_{1,m} .  \label{eq:swap}
\end{eqnarray}
Consider now the reduced system dynamics defined by
\begin{eqnarray}
\label{new}
\!\!  \rho_{S}' (n) & {=} & \tmop{Tr}_{E_{n} \ldots E_{1}} \mathcal{U}_{SE_{1}}
  \mathcal{S_{\nosymbol}}_{1,n} \mathcal{V}_{E_{n} E_{1}}  \mathcal{U}_{SE_{1}}
  \ldots \\
  &  & \ldots \mathcal{S_{\nosymbol}}_{1,3} \mathcal{V}_{E_{3} E_{1}} 
  \mathcal{U}_{SE_{1}}  \mathcal{S}_{1,2}  \mathcal{V}_{E_{2} E_{1}} 
  \mathcal{U}_{SE_{1}}  \!\!\left[ \rho_{S} \otimes \bigotimes_{i=1}^{n}
  \rho_{E_{i}} \right] \,, \nonumber
\end{eqnarray}
where the initial state is again assumed to be given by Eq.~\eqref{initial}. In such a dynamics, the first ancilla $E_1$ works as a 
``memory" in the following sense: The system begins by colliding with $E_1$ (memory) according to the map $\mathcal U$. The memory then collides with the next ancilla $E_2$ via map $\mathcal{V}$, which is followed by a swap $\mathcal{S}$ on $E_1$ and $E_2$, as shown in Fig.~\ref{FigBassano1} {\bf (b)}. The sequence is then iterated, so that the system at each step directly interacts with the memory $E_1$ only, while $E_1$ collides successively with the remaining ancillas one at a time according to the effective collision map $\mathcal S_{1,n}\circ\mathcal{V}_{E_nE_1}$. We next prove that the open dynamics defined by \eq\eqref{new} coincides with the dynamical map \eqref{eq:scatola2}.

Using Eq.~(\ref{eq:swap}), we can move all swap operators to the left in \eq\eqref{new} and update indexes accordingly as
\begin{eqnarray}
  \rho_{S}' (n) & = & \tmop{Tr}_{E_{n} \ldots E_{1}} \mathcal{S}_{1,n} 
  \mathcal{U}_{SE_{n}}  \mathcal{\mathcal{S}}_{1,n-1}  \mathcal{V}_{E_{n}
  E_{n-1}}  \mathcal{U}_{SE_{n-1}} \ldots \nonumber\\
  &  & \ldots \mathcal{\mathcal{S}}_{1,2}  \mathcal{V}_{E_{3} E_{2}} 
  \mathcal{U}_{SE_{2}}  \mathcal{V}_{E_{2} E_{1}}  \mathcal{U}_{SE_{1}} 
  \left[ \rho_{S} \otimes \bigotimes_{i=1}^{n} \rho_{E_{i}} \right] .
  \nonumber
\end{eqnarray}
We next move each partial trace $\tmop{Tr}_{E_{m}}$ for $2\le m\le n-1$ to the right until it meets a map acting on $E_m$, thus obtaining
\begin{eqnarray}
  \rho_{S}' (n) & = & \tmop{Tr}_{E_{n} E_{1}}\! \mathcal{S}_{1,n} 
  \mathcal{U}_{SE_{n}} \!\tmop{Tr}_{E_{n-1}} \mathcal{S}_{1,n-1} 
  \mathcal{V}_{E_{n} E_{n-1}} \! \mathcal{U}_{SE_{n-1}} \ldots \nonumber\\
  &  & \ldots \tmop{Tr}_{E_{2}}\! \mathcal{S}_{1,2}  \mathcal{V}_{E_{3} E_{2}} 
  \mathcal{U}_{SE_{2}}  \mathcal{V}_{E_{2} E_{1}}  \mathcal{U}_{SE_{1}}\!\! 
  \left[ \rho_{S} \otimes \bigotimes_{i=1}^{n} \rho_{E_{i}} \right]\!.
  \nonumber
\end{eqnarray}
This expression suggests that indeed to match the exact reduced dynamics at most correlations within a three-qubit system needs to be taken into account. With the help of \eq\eqref{TrS} we can now recursively get rid of the swap operations so as to end up with
\begin{eqnarray}
  \rho_{S}' (n) & {=} &\! \tmop{Tr}_{E_{n}}\! \mathcal{U}_{SE_{n}}
 \! \!\tmop{Tr}_{E_{n-1}}\! \mathcal{V}_{E_{n} E_{n-1}}  \mathcal{U}_{SE_{n-1}}
  \ldots \nonumber\\
  &  & \ldots \tmop{Tr}_{E_{2}} \!\!\mathcal{V}_{E_{3} E_{2}} 
  \mathcal{U}_{SE_{2}} \tmop{Tr}_{E_{1}}\!\! \mathcal{V}_{E_{2} E_{1}} 
  \mathcal{U}_{SE_{1}}\!\!  \left[ \rho_{S} {\otimes} \bigotimes_{i=1}^{n}
  \rho_{E_{i}} \!\right] \nonumber\\
  & = & \tmop{Tr}_{E_{n} \ldots E_{1}} \mathcal{U}_{SE_{n}} \ldots
  \mathcal{V}_{E_{2} E_{1}}  \mathcal{U}_{SE_{1}}  \left[ \rho_{S} \otimes
  \bigotimes_{i=1}^{n} \rho_{E_{i}} \right] , \nonumber
\end{eqnarray}
which coincides with \eq\eqref{eq:scatola2} thus identifying $\rho_{S}' (n)$ and $\rho_{S} (n)$. This concludes the proof. 

The dynamical map $\Lambda (n)$ in Eq.~(\ref{new}), which describes the open dynamics of $S$, thus admits the equivalent representation (\ref{eq:l1}). Note that the above equivalence holds regardless of the form of all the involved collision maps, the initial states of the ancillas, as well as the dimensionality of $S$ and ancillas.

As discussed, considering interactions involving a larger number of ancillas (i.e., $d>1$) stronger memory effects can be featured in the framework of collision models~\cite{BarisPRA}. This situation can also be reformulated considering a memory of higher dimensionality, whose size is associated to the memory depth, acting as mediator between system and environment. To this aim let us come back to the evolution described by Eq.~(\ref{eq:3b}). In this case one can consider an equivalent dynamics such that $S$ is repeatedly interacting with {\it two} environmental ancillas, which mediate the coupling to the environment by undergoing collisions with ancillas. Indeed, in analogy with \eq\eqref{new}, the dynamics defined by 
\begin{widetext}
\begin{equation}
 \rho_{S}^{'''} ( n )  =  \tmop{Tr}_{E_{2n} \ldots E_{1}}
  \mathcal{U}_{SE_{2} } \mathcal{U}_{S E_{1}} \mathcal{\mathcal{S}}_{2,2n}
  \mathcal{\mathcal{S}}_{1,2n-1} \mathcal{V}_{E_{2n-1} E_{2} E_{1}}
  \ldots \mathcal{\mathcal{S}}_{2,6} \mathcal{\mathcal{S}}_{1,5} \mathcal{V}_{E_{5} E_{2} E_{1}}  \mathcal{U}_{SE_{2}} \mathcal{U}_{SE_{1}} \mathcal{\mathcal{S}}_{2,4} \mathcal{\mathcal{S}}_{1,3} 
   \mathcal{V}_{E_{3} E_{2} E_{1}}  \mathcal{U}_{SE_{2}} \mathcal{U}_{SE_{1}} \left[ \rho_{S}
  \otimes \bigotimes_{i=1}^{2n} \rho_{E_{i}} \right] ,
  \label{eq:33bis}
\end{equation}
\end{widetext}
can be shown to be equivalent to Eq.~(\ref{eq:3b}). Therefore, also for {\it ``second-order-Markovian''} dynamics described by the maps $\{ \Lambda^{(2)} (n) \}$ one can consider an equivalent representation of the dynamics by means of a memory whose size is determined by the memory depth. A similar approach can be used for arbitrary $d$. At this point we remark on the requirement for $S$ to interact with the whole memory before the intra-environment collisions take place. If in Eq.~\eqref{Un}, $d$-ranged AA collisions occurred after every {\it single} $S$-$E_i$ collision, it would remain true that the memory depth, i.e. the relevant correlations to the dynamics, would reside within the last $d$ ancillas the system interacted with. However, such a setting does not allow us to exploit the swap operation to ensure the system only interacts with the {\it same} ancillas throughout its entire dynamics.

\section{\mbox{Relevant Correlations: Thermodynamics}}\label{thermodynamics}
\subsection{Entropy}
The previous two sections showed that if one aims at describing the, in general, non-Markovian {\it open system dynamics}, then only correlations between the system and a bath portion as large as the range of intra-environment interactions needs to be accounted for. One may wonder if this or a suitably adapted property holds in the characterization of the {\it thermodynamical} features, a task for which CMs are increasingly used~\cite{EspositoPRX}. 

For the sake of argument, we will refer to a CM featuring only nearest-neighbor AA collisions, corresponding to a {\it ``first-order-Markovian''} process as described by Fig.~\ref{fig1}, Eq.~\eqref{eq:l1}, and explicitly considered in Sec.~\ref{erasure}. We will focus on the behavior of entropic quantifiers related to irreversible entropy production (as explained shortly) and, in the last part, heat flux.

Let $S_{S}$ be the von Neumann entropy of the system, $S_S=-{\rm Tr}(\rho_S \log \rho_S)$. Exploiting the properties of the relative entropy, similarly as in \rref\cite{EspositoNJP}, the dynamical change of $S_S$ can be  expressed in terms of system-environment correlations, changes in the environment's state as well as heat exchanged between system and environment. Indeed, after some manipulation, the change in the von Neumann entropy of $S$, 
\begin{align}
\delta S_{S} ( n ) &{ =}  S_{S} ( n ) -S_{S} ( 0 ) \label{DeltaS}
\end{align}
can be exactly decomposed as
\begin{eqnarray}
  \delta S_{S} ( n ) & = &
   \underbrace{I_{SE} ( n ) -I_{SE} ( 0
  )}_{\mathcal{S}_{\tmop{corr}}} \nonumber \\ 
  & & + \underbrace{S ( \rho_{E} ( n ) |
  \nobracket \rho_{E} ( 0 ) ) -S ( \rho_{E} ( 0 ) | \nobracket \rho_{E} ( 0
  ) )}_{\mathcal{S_{\tmop{env}}}} \nonumber\\
  &  & + \underbrace{\tmop{Tr}_{E} ( \rho_{E} ( n ) - \rho_{E} ( 0 ) ) \log
  \rho_{E} ( 0 )}_{\mathcal{Q}} \nonumber\\
  & = & \mathcal{S}_{\tmop{corr}} +\mathcal{S}_{\tmop{env}} + \mathcal{Q}
  \label{BassanoEq}
\end{eqnarray}
where $S(\rho|\sigma)={\rm Tr}\rho(\log \rho-\log \sigma)$ is the relative entropy. In particular, for $\rho_{E} ( 0 ) = e^{-\beta \hat H_E}/{{\rm Tr}}(e^{-\beta \hat H_E})$, i.e. a Gibbs state, the third heat-like term is given by
\begin{eqnarray}
\mathcal{Q}& = &  - \beta ( \langle \hat H_{E} \rangle_n-
  \langle \hat H_{E} \rangle_0). \label{Qex}
\end{eqnarray}
In the standard separation of terms of entropy production \cite{Mazur}, which calls for a connection to a thermodynamic viewpoint, the contribution $\mathcal{S}_{\tmop{corr}} +\mathcal{S}_{\tmop{env}}$ is usually interpreted as the {\it irreversible} entropy production. As shown by \eq\eqref{BassanoEq}, this is related to the establishment of system-environment correlations and changes in the environmental state~\cite{EspositoNJP}. The third contribution in \eq\eqref{BassanoEq} embodies instead the reversible entropy production, associated to the heat exchanged with the environment [\cf\eq\eqref{Qex}]. Note that the sum of the last two contributions can also be seen as (minus) the variation of the environment entropy over the given time interval, namely $\delta S_{E} ( n )$.
\begin{figure}[t]
{\bf (a)} \\ 
\includegraphics[width=0.75\columnwidth]{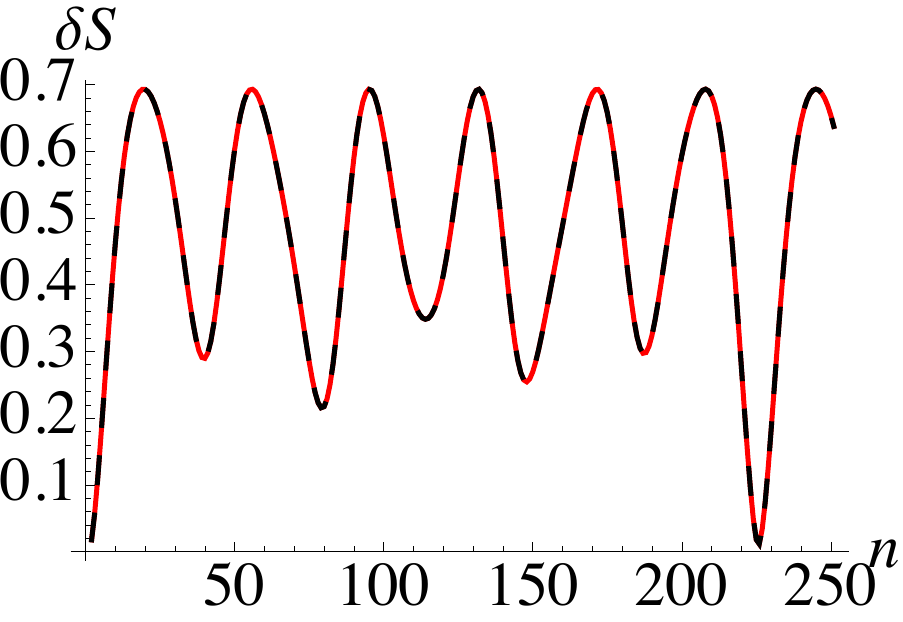}\\
{\bf (b)} \\
\includegraphics[width=0.75\columnwidth]{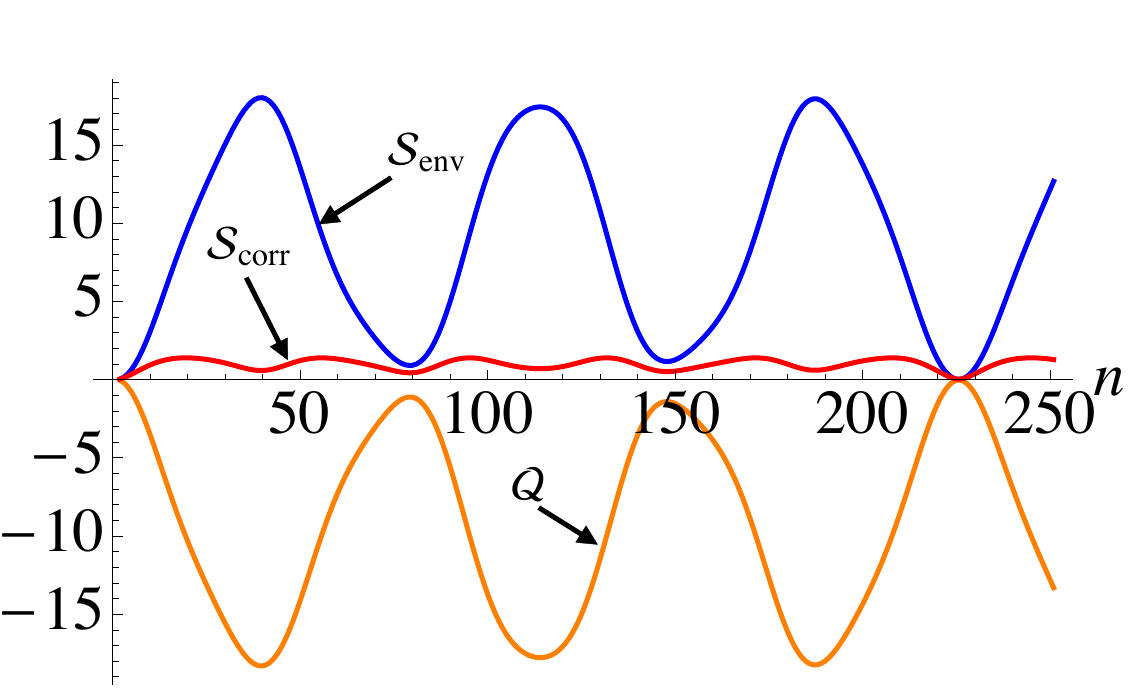}
\caption{{\bf (a)} Simplified CM featuring only two ancillas (with which $S$ interacts successively and iteratively): change in entropy, $\delta S_S$ against the number of steps $n$, determined using Eq.~\eqref{DeltaS} (solid, colored) and Eq.~\eqref{BassanoEq} (dashed, black). We fix the $S$ initial state to be $\ket{1}$ and the ancillas in $\ket{0}$. {\bf (b)} We plot the three different contributions entering into Eq.~\eqref{BassanoEq}. We assume all collisions are partial-SWAPs with weak $S$-$E_n$ collisions, $J\tau_{SA}=0.05$, and strong AA collisions with $J\tau_{AA}=0.95\tfrac{\pi}{2}$.}
\label{fig5}
\end{figure}

To illustrate the behaviour of the three entropic terms in \eq\eqref{BassanoEq} and how they combine to give the entropy change of $S$ \eq\eqref{DeltaS}, let us first consider a greatly simplified collision model, where the environment consists of only two qubits which the system collides with sequentially. Thus, the situation is similar to Fig.~\ref{fig1} if we restrict the environment to consist of only ancillas $E_1$ and $E_2$ which $S$ {\it iteratively} interacts with, neither of which is ever traced over nor any correlations shared between the three qubits is ever discarded. In Fig.~\ref{fig5} {\bf (a)} we show the change in entropy evaluated using Eqs.~\eqref{DeltaS} and \eqref{BassanoEq}, where we have arbitrarily fixed the initial state of the environment ancillas to be in their respective ground states and assume the partial-SWAP operation for the collision interactions. The three components of Eq.~\eqref{BassanoEq} are plotted separately in panel {\bf (b)}, showing in particular that $\mathcal{S}_\text{corr}$ and $\mathcal{S}_\text{env}$ are strictly positive, in line with their association to the irreversible entropy production. As expected, the sum of $\mathcal{S}_\text{corr}(n)$, $\mathcal{S}_\text{env}(n)$ and ${\cal Q}(n)$ matches $\delta S_S(n)$.
\begin{figure}[t]
	{\bf (a)} \\ 
	\includegraphics[width=0.75\columnwidth]{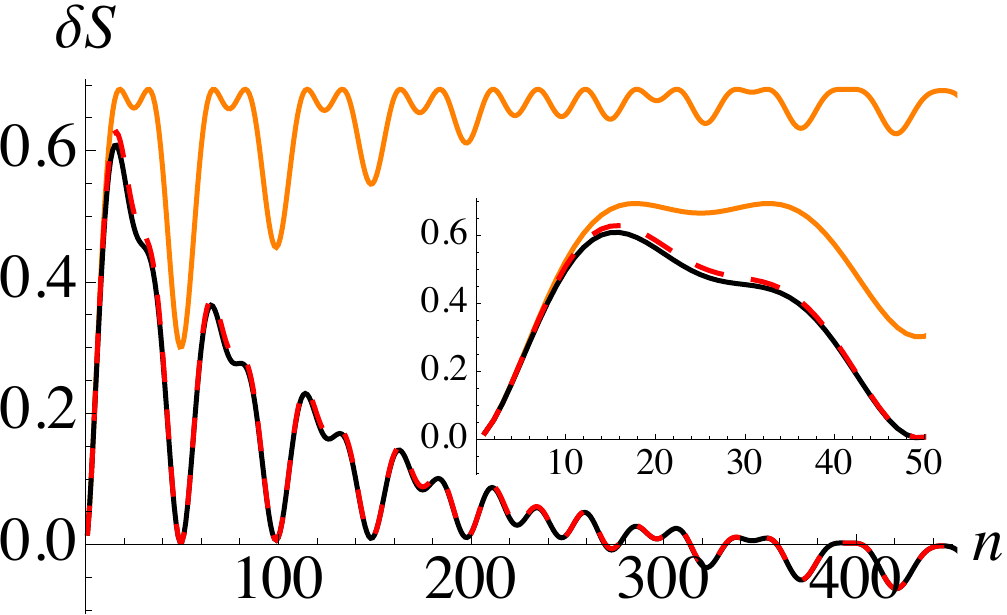}\\
	{\bf (b)} \\
	\includegraphics[width=0.75\columnwidth]{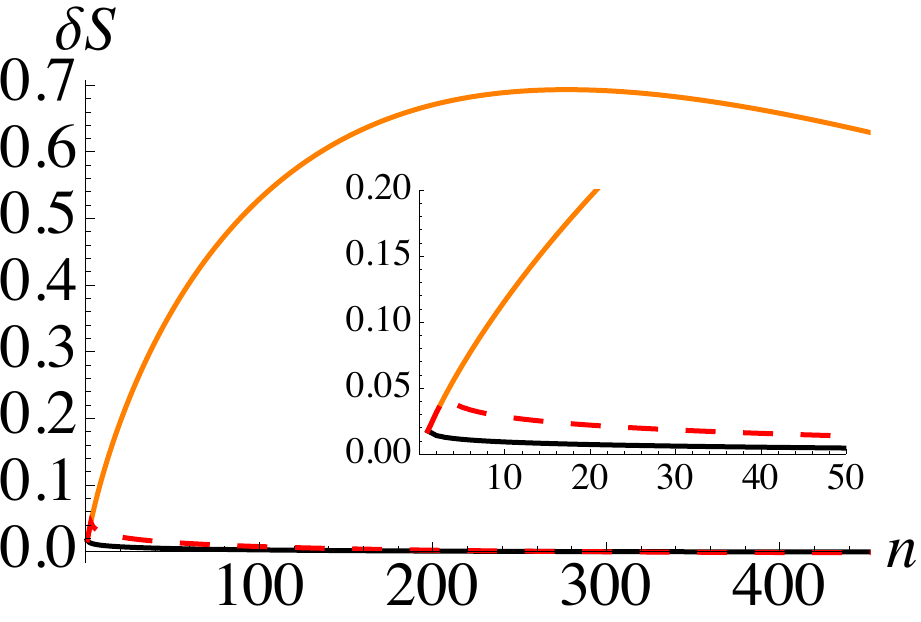}
	\caption{CM with nearest-neighbor AA collisions: change in entropy $\delta S_S(n)$ evaluated using Eq.~\eqref{DeltaS} (top-most, orange), and using Eq.~\eqref{BassanoEq} for erasure schemes {\bf B} (bottom-most solid black) and {\bf C} (dashed, red) in the case of strong AA collisions with $J\tau_{AA}=0.95\tfrac{\pi}{2}$ {\bf (a)} and in the absence of AA collisions {\bf (b)} . We assume the system is initially in its excited state $\ket{1}$ and the environmental subunits are initially in their ground states. Throughout we assume all collisions are partial-SWAPs with weak $S$-$E_n$ collisions, $J \tau_{SA}=0.05$. Insets show the behaviour in the first 50 steps.}
	\label{fig8}
\end{figure}

Now consider the non-Markovian CM of Fig.~\ref{fig1} with nearest-neighbor AA collisions which, as exhaustively demonstrated in Secs.~\ref{erasure} and \ref{depth}, showed that retaining correlations between $S$ and two ancillas at each step is enough to fully capture the open dynamics and, in particular, the degree of non-Markovianity. In light of Fig.~\ref{fig5} one can thus wonder whether by replacing the entire environment with only the few relevant ancillas to the dynamics at each step in \eq\eqref{BassanoEq}, the system's entropy change \eq\eqref{DeltaS} is again retrieved. This {\it does not} occur (not even approximately) as shown in \fig\ref{fig8} {\bf (a)} where we compare the behavior of $\delta S_S(n)$ and \eqref{BassanoEq} for the erasure scheme {\bf B} (lower black), where two ancillas and their correlations are stored at each step, and {\bf C} (dashed red) where three ancillas are retained at each step. Clearly in the very short-term dynamics the curves align closely, however large discrepancies quickly emerge. This discrepancy is not restricted to non-Markovian environments, but occurs even if memory effects are fully absent. This can be seen from \fig\ref{fig8} {\bf (b)} where AA collisions are switched off, so as to retrieve a fully Markovian situation, showing again a large discrepancy between \eqs\eqref{DeltaS} and \eqref{BassanoEq}. Thus, it is clear that scheme {\bf A} also exhibits a large discrepancy qualitatively similar to Fig.~\ref{fig8} (results not shown).

These results strongly indicate that, while only correlations between the system and a portion of the bath are relevant for the open dynamics, the description of thermodynamical properties generally demands to account for correlations with the {\it entire} environment.

\subsection{Heat flux and non-Markovianity}
For other thermodynamic quantities, exceptions to the above general framework may occur for energy-preserving system-environment couplings \cite{RuariPRL, LorenzoPRA2015}, for instance the Heisenberg interaction [i.e., for $J_x\!=J_y\!=J_z$ in \eq\eqref{HI}]. In such cases, the heat exchanged by the system fulfils $\mathcal{Q}_S=-\mathcal{Q}_E$, where $\mathcal{Q}_S = \text{Tr}\left[\hat H_S(\rho_S(0) {-}  \rho_S(n))  \right]$ while
\begin{equation}
\label{heatE}
\mathcal{Q}_E = \sum_n \text{Tr}\left[ \hat H_E \left( \rho_{E_{n}} - \tilde{\rho}_{E_{n}} \right) \right]\,,
\end{equation} 
where $\rho_{E_{n}}$ ($\tilde{\rho}_{E_{n}}$) is the state of the $n$th ancilla before (after) it has interacted with the system. 

\begin{figure}[t]
\includegraphics[width=0.85\columnwidth]{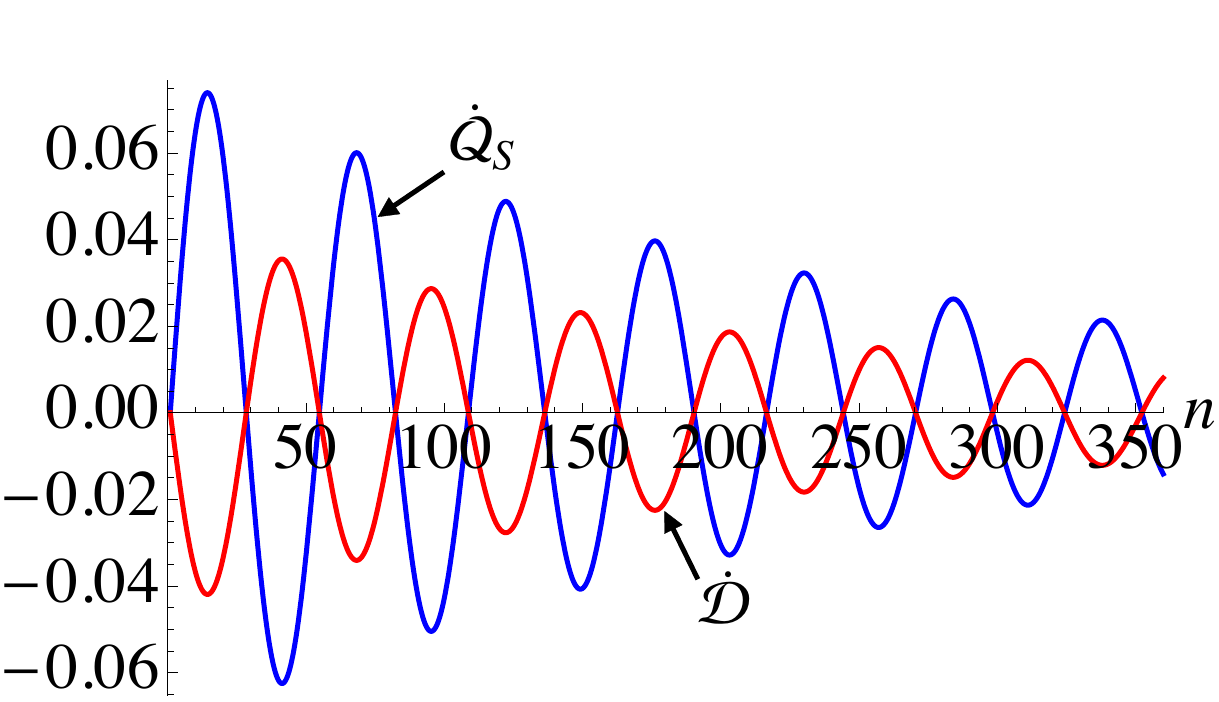}
\caption{Behavior of heat flux $\dot{\mathcal{Q}}_S$ (blue line) and time-derivative of the trace distance between the evolution for initial systems states $\{\ket{0},\ket{1}\}$ (red). We consider a CM with nearest-neighbor AA partial-SWAP collisions corresponding to $J \tau_{AA}=0.95\tfrac{\pi}{2}$ and weak system-ancilla collisions such that $J \tau_{SA}=0.05$. We assume each ancilla to be initially in a Gibbs state with inverse temperature $\beta=1$.}
\label{fig7}
\end{figure}

The behavior in time of the heat flux, $\dot{\mathcal{Q}}_S$, is generally non-monotonic if memory effects are present \cite{RuariPRL, LorenzoPRA2015} as shown in a paradigmatic instance in Fig.~\ref{fig7}. In the same figure, we plot the behavior of the trace distance for the initial pair of states of $S$ $\{\ket{0},\ket{1}\}$. In this instance a relationship emerges between heat flux and the trace distance: their behaviors are perfectly aligned with one-another, i.e., a non-monotonic behavior of the heat flux is commensurate with the on-set of a non-Markovian dynamics (this can be found with any choice of initial system states except the steady state). 

This feature provides a thermodynamic interpretation of non-Markovianity as quantified by information back flow \cite{BreuerPRL} in the case of energy-preserving system-environment interactions. This is because in such cases information flow between the system and environment is always accompanied by energy exchange.

\section{Conclusions}
\label{conclusions}
In this work, through a collision-model-based approach, we investigated the relevance of system-environment correlations in connection with the possibility to embed the system's non-Markovian dynamics into a Markovian one once the system is extended so as to include a suitably sized bath portion. We considered a simple collision model where the bath is made out of a large collection of ancillas, which the system successively collides with. A memory mechanism is enabled by the occurrence of collisions between the environmental ancillas so as to make the system dynamics generally non-Markovian. Such models allow to study the relevance of correlations by analyzing how the dynamics is affected by the erasure of correlations between the system and environmental ancillas. We reviewed and generalized the results of \rref\cite{RuariPRA}, showing that only correlations between the system and a bath portion as large as the range of intra-environmental interactions matter to the open dynamics, in particular the degree of non-Markovianity. Building on this, we presented a general framework supporting an effective description of the open dynamics. This shows, in particular, how to construct a Markovian embedding for the system dynamics depending on the range of intra-environmental interactions by introducing the notion of memory depth. Additionally, we provided evidence that, at variance with open system dynamics, even in a fully Markovian situation and irrespective of the range of intra-environmental interactions, {\it all} system-environment correlations are generally relevant to the description of thermodynamical quantities such as entropy production. Exceptions can however occur for energy-preserving system-environment interactions, in which case we showed that one can give a thermodynamical interpretation in terms of heat flux of the well-known non-Markovianity indicator based on information back flow \cite{BreuerPRL}.

As our analysis has focussed on quantum dynamics described by a certain class of CMs, we finally address the generality of our conclusions. While any Markovian (CP-divisible) open dynamics can be reproduced through a suitably defined CM, it is not as yet known whether the analogous property holds for arbitrary non-Markovian dynamics, although it was shown this is the case in some paradigmatic instances~\cite{StrunzPRA2016, lorenzo2017a, lorenzo2017b}. Much of the spirit of our study comes from quantum thermodynamics, where CMs are becoming a popular tool to answer conceptual questions that require some knowledge of the bath dynamics. The class of non-Markovan CMs considered here is conceptually significant in that it enjoys at once two usually demanding properties~\cite{CiccarelloPRA2013, VacchiniPRA2013}: it leads to a reduced master equation for the system that, like the Lindblad master equation, is ensured to be completely positive. However, in sharp contrast to the Lindblad master equation, this master equation is able to capture strong non-Markovian behaviour. The work presented here, in particular Sec.~\ref{depth}, thus serves as a significant case study. Its formal extension, and in particular of the intuitive notion of memory depth, to more general non-Markovian dynamics, starting from non-Markovian CMs that rely on different memory mechanisms, is a task for future investigations.

\acknowledgements
We acknowledge support from the EU Collaborative project QuProCS (grant agreement 641277) and FFABR.

\bibliography{collision_models}

\end{document}